# STRUCTURAL ORIGIN OF THE METAL-INSULATOR TRANSITION OF MULTIFERROIC BiFeO$_3$

S. A. T. Redfern[1], J. N. Walsh[1], S. M. Clark[2], G. Catalan[1], and J. F. Scott[1]

[1]Department of Earth Sciences, University of Cambridge, Downing Street, Cambridge CB2 3EQ, UK

[2]Advanced Light Source, Lawrence Berkeley National Laboratory, 1 Cyclotron Road, Berkeley, CA 94720-8226, USA

## Abstract

We report X-ray structural studies of the metal-insulator phase transition in bismuth ferrite, BiFeO$_3$, both as a function of temperature and of pressure (931 $^o$C at atmospheric pressure and ca. 45 GPa at ambient temperature). Based on the experimental results, we argue that the metallic γ-phase is not rhombohedral but is instead the same cubic Pm3m structure whether obtained via high temperature or high pressure, that the MI transition is second order or very nearly so, that this is a band-type transition due to semi-metal band overlap in the cubic phase and not a Mott transition, and that it is primarily structural and not an S=5/2 to S=1/2 high-spin/low-spin electronic transition. Our data are compatible with the orthorhombic Pbnm structure for the β-phase determined definitively by the neutron scattering study of Arnold et al .[Phys. Rev. Lett. 2009]; the details of this β-phase had also been controversial, with a remarkable collection of five crystal classes (cubic, tetragonal, orthorhombic, monoclinic, and rhombohedral!) all claimed in recent publications.

Bismuth ferrite BiFeO$_3$ (BFO) has become the cornerstone of research in magnetoelectric multiferroics [1, 2] due to it being a rare (though perhaps not unique [4, 3]) example of a simple perovskite oxide with strong ferroelectric polarization [5, 6] and magnetic ordering [7, 8, 9] at room temperature, with interesting coupling between the two ferroic orderings [10, 11]. Yet in spite of all the attention that it has received, several basic aspects still remain unresolved. One of these is the phase diagram, which is turning out to be rather complex, with several new phase transitions being reported just in the last year [12-20]. Bismuth ferrite is simultaneously ferroelectric, antiferromagnetic and ferroelastic, that is, there are at least three ferroic order parameters involved in its phase transitions, and other crystallographic distortions, such as rotations of the oxygen octahedra, also play an important role in the functional properties. Given the number of order parameters involved, and the subtle coupling between them, it is perhaps not unreasonable that the phase diagram should be so rich.

Here we would like to focus our attention on the high-pressure and high-temperature ends of the phase diagram, where a metal-insulator (MI) transition is known to occur [12, 13, 21, 22]. Optical and transport studies have previously shown that BFO becomes metallic at 930 $^o$C and ambient pressure [12], or at ca. 45 GPa and room temperature [13, 22, 23]; however, the nature of this MI transition and the associated structural changes is still not clear, and conflicting models have been proposed. Specifically, it is not at this point clear whether the MI transition is of band-type [12] or Mott type [13]. A band-type insulator has an even number of electrons in the unit cell, and these fill completely the valence band, so that there is a gap between them and the excited states in the conduction band. For a valence band insulator to become a metal, there has to be a structural transition whereby the number of formula units (and therefore the number of electrons) per unit cell changes. In a Mott-type insulator, by contrast, the gap is due to electrostatic repulsion between the conduction electrons. A transition from Mott-insulator to metal is thus one in which the size of the electrostatic repulsion (the Mott-Hubbard parameter U) becomes smaller than the width of the conduction band [24, 25]. Mott transitions are purely electronic and do not in theory require a change in either crystal structure or magnetic symmetry, although in practice the coupling between charge, spin and lattice means that other transitions tend to happen simultaneously [26]. The challenge, often, is to find which comes first: does the structural transition drive the electronic one (band MI transitions) or is it the other way round (Mott MI transitions).

Based on diffraction experiments, it is argued here that the nature of the MI transition in BFO is the same irrespective of whether it is achieved with temperature or with pressure, and is primarily due to a structural change from orthorhombic to cubic symmetry. The key structural parameter is identified as the rotation of the oxygen octahedra, which disappears in the cubic phase thereby enhancing the orbital overlap between oxygen and iron ions.

Powdered BiFeO$_3$ was studied between room temperature and 1000$^o$C using a Bruker D8 Advance X-ray diffractometer in $\theta-\theta$ geometry. Diffraction patterns were collected between 20 and 90$^o$ $\theta$ CuK$\alpha$ radiation using a rapid Vantec position sensitive detector. This allowed patterns to be obtained on a time scale of less than 10 minutes, with rapid scanning rates. Such rapid data collection is absolutely essential at the highest temperatures of the experiment since BiFeO$_3$ is not chemically stable in air or vacuum at such temperatures, and upon entering the cubic phase breaks down as a function of time. However, with this precaution of rapid data collection, the samples were cycled 5 times each to >931°C without decomposition. We note that phase transitions to higher symmetry structures often trigger dissociation due to the higher entropy (e.g., the breakdown of calcite on entering its high-T phase– [27] or the melting of ferroelectric LiNbO$_3$ on entering the paraelectric phase [28]).

Figure 1 illustrates the resolution of the three most intense diffraction lines in the $2\theta$-plot of XRD results, including the strongest (110)$_c$ line (subscript "c" here refers to indexing based on the primitive cubic unit cell). We display XRD data as a function of *d*, since different X-ray excitation wavelengths were used in the high-T and high-P experiments. Above ca. 1200K, there is a single peak, compared with a large splitting in the rhombohedral phase at ambient temperatures (a small asymmetry arises from $\alpha_1/\alpha_2$ source wavelengths). This highest temperature phase (γ phase) therefore appears to be cubic. Below 1200K there is a transition to a β phase. The lattice constant data published earlier [12] indicated that the β-phase is orthorhombic and not tetragonal (nor pseudo-tetragonal), and that the β-γ (orthorhombic-cubic) phase transition is continuous or very nearly so. Very recent neutron powder diffraction studies at high temperature have confirmed the orthorhombic structure of the β-phase below the MI transition [15].

The high pressure data were collected on BiFeO$_3$ loaded into a diamond anvil cell and pressurised to around 50 GPa, pressures measured by the fluorescence of ruby chips in close proximity to the sample, using the high-pressure beam line 12.2.2 of the Advanced Light Source, Lawrence Berkeley Laboratory. Data were collected onto Marr image plates and converted into one dimensional diffraction patterns using standard methods, by Fit2D. Fig. 2 illustrates the most intense [110] XRD line as a function of pressure (to 47 GPa). The same cubic/non-cubic transition seems evident. We note parenthetically that we clearly see the low-pressure phase transitions near 5-10 kbar recently reported by Haumont *et al.* [17], both as a change in the diffraction pattern and as a slight contraction of the unit-cell volume near 7.5 GPa (see Figure 3), which is similar to, though smaller in magnitude than, the compression reported for the α-β (rhombohedral-orthorhombic) transition as a function of temperature [12]. The evolution of the unit cell volume as a function of pressure is compared in figure 3 with previously reported results from Gavriliuk et al, [13, 23]. Both sets of data agree quantitatively rather well. Importantly also, when plotted together, the two appear consistent with a continuous evolution and no sharp changes in unit cell volume, suggesting a second order phase transition.

In order to consider whether the cubic phase we infer at high pressure is the same as that which we observe at high temperature, we show in Fig.4 a wider range of XRD data. The high-pressure data suffer from broadening due to lower hydrostaticity at extreme pressure, plus obviously the lattice parameter is considerably smaller, but the patterns at both high-pressure and high-temperature are identical in relative peak positions and heights, and are both compatible with a primitive cubic structure in both cases. We emphasize also the disappearance of the superlattice peaks characteristic of the orthorhombic β phase. The disappearance of all superlattice peaks indicates that the diffraction pattern corresponds to a simple perovskite unit cell (i.e., there is no unit cell doubling), indicating clearly that the high pressure phase cannot be orthorhombic Pbnm, nor rhombohedral R3c, as both of these have unit cells that are multiples of the primitive perovskite. This is also important regarding the MI transition: simple cubic BFO has only one formula unit per unit cell, with an odd number of valence electrons

(there are 5 electrons in the d-shell of $Fe^{+3}$), whereas orthorhombic and rhombohedral $BiFeO_3$ possess 2 formula units per unit cell, with an even number of electrons; accordingly, on cooling from cubic to orthorhombic BFO can become a band insulator.

The results therefore indicate the same sequence of phase transitions as a function of increasing temperature or increasing pressure. The first structural phase transition, α-β (rhombohedral to orthorhombic) is first order, as indicated by sharp volume contraction. In the vicinity of this first order phase transition there can be phase coexistence of the α and β phases [13], which may have contributed to the past discrepancies about the nature of this $\beta$ phase [14, 16, 29, 30]. The variety of crystal classes wrongly attributed to this phase is itself also remarkable, as the list includes cubic [29], tetragonal [30], monoclinic [16] and rhombohedral [14]. While this may seem surprising, it is less so when put in the context of i) the difficulty in the interpretation of the patterns due to phase coexistence, ii) the very high temperatures at which the measurements are done, which contribute to sample decomposition and peak broadening and iii) the low sensitivity of x-ray diffraction to oxygen positions, which are key.

Contrary to the α-β transition, the β-γ phase transition at high temperature or high pressure appears to be essentially continuous (second order), which is significant because second order MI transitions cannot be Mott-type [24]. As a function of increasing temperature or pressure, a decrease in optical bandgap and resistivity has been reported [12, 12]. Upon entering the orthorhombic β-phase, the resistivity decreases further, but BFO remains still semiconducting [12, 14]. When the cubic phase is finally reached, BFO becomes metallic [12]. The evidence thus suggests that the bandgap is directly linked to the crystallographic distortion, and that the structural change may be sufficient to drive the metal-insulator transition, rather than the other way round.

In earlier high pressure studies, however, a different scenario was proposed. It was noted that the MI transition coincides with a change in the magnetic configuration of the $Fe^{3+}$ ions from high spin to low spin [13], a finding also supported by first principles calculations [18]. Gavriliuk *et al.* hence proposed [13] that strong electron-electron repulsion (the Mott-Hubbard parameter U) in the high-spin phase could be responsible for the opening of the bandgap that causes BFO to be an insulator. In the low-spin phase this electrostatic repulsion would be smaller, enabling the bandgap to decrease, leading to a metallic state. Gavriliuk *et al.* also mention the presence of Mott's variable range hopping [32] in the semiconducting phase as consistent with this, although variable range hopping is not itself a proof of a Mott-type phase transition, nor is it likely to exist at room temperature [33]: it is unphysical to consider tunnelling over a length scale that is larger than the inelastic mean free path [34], which is of the order of a unit cell at room temperature [35].

Several key aspects of Gavriliuk's model --the existence of a high-spin to low-spin transition at high pressure, the weakness of the magnetic interactions in the low-spin phase leading to paramagnetism at room temperature, and the existence of a sizeable density of states at the fermi level (i.e., metallicity) in a paramagnetic low spin phase-- are also supported by ab-initio calculations [18]. On the other hand, the first principle calculations [12, 18] also show the valence bands to be too broad and strongly hybridized to be compatible with a Mott-Hubbard origin of the gap. There are also other points that weight in favour of a band-type and against a Mott-type phase transition: for example, band-structure calculations show that cubic BFO is a semimetal [12], which implies that a structural phase transition to a cubic phase is by itself enough to cause a band-type insulator-metal transition. Also Mott's requirement that the MI transition be first order is at odds with the observed continuous evolution of the lattice parameter/unit cell volume. The Mott-type MI transition is in fact defined as an isostructural transition from paramagnetic metal to paramagnetic insulator transition [24-26]. This is not the case in BiFeO$_3$, where both a structural transition (from orthorhombic to cubic symmetry), and a magnetic phase transition (from antiferromagnet to paramagnet [21]) take place simultaneously at high pressure. Finally, The MI transition at high temperature is not due to a change in spin ordering (BFO is paramagnetic in both phases), so the high-spin to low-spin model of the transition does not apply to it; conversely, since the structural changes are the same as those as a function of pressure, the MI transition mechanism is likely to also be the same.

We therefore believe that both the temperature-driven and the pressure-driven MI transitions are controlled by the crystallographic change and not by the electron-electron repulsion. The key structural parameter is likely to be the straightening of the Fe-O-Fe bond angle, which leads to increased orbital overlap between the oxygen *p* orbital and the iron *d* orbital thereby facilitating charge transfer. This angle is known to play a key role in the functional properties of transition metal perovskite oxides [37], and its effect on BFO is backed by experimetal results: as shown in Figure 5, there is a clear inverse correlation between the evolution of the Fe-O-Fe bond angle and the bandgap. Using a crude linear extrapolation from the rhombohedral phase, we estimate that a critical angle of ca. ~159$^\circ$ degrees may be enough to eliminate the bandgap altogether and trigger a metallic state at high temperature, even in the absence of a structural phase transition. This critical angle is in fact amply surpassed in the cubic phase, where it is 180$^\circ$. The correlation between octahedral rotation and bandgap in BiFeO$_3$ is also supported by recent studies of the ferroelectric domain walls, whose local structure is characterized by decrease in octahedral rotation leading to decreased bandgap and increased conductivity [38]. Finally, it is also worth mentioning that cubic SrFeO$_3$ is also a metal and, although the Fe$^{+4}$ oxidation state in this compound precludes direct comparison with BFO, we think that the structural correlation is not a coincidence.

In summary, the diffraction results show the same orthorhombic-cubic phase transition at high temperatures (930 $^\circ$C) or pressure (~45GPa), and support the view that the metal-insulator transition in BFO is a conventional band-type transition triggered by the symmetry change, with the key structural parameter being the Fe-O-Fe bond angle.

*Acknowledgements.* We thank John Robertson, Jorge Iñiguez and Peter Littlewood for useful discussions regarding this manuscript. The financial support of the EC-STREP program MULTICERAL is also gratefully acknowledged.

Figures:

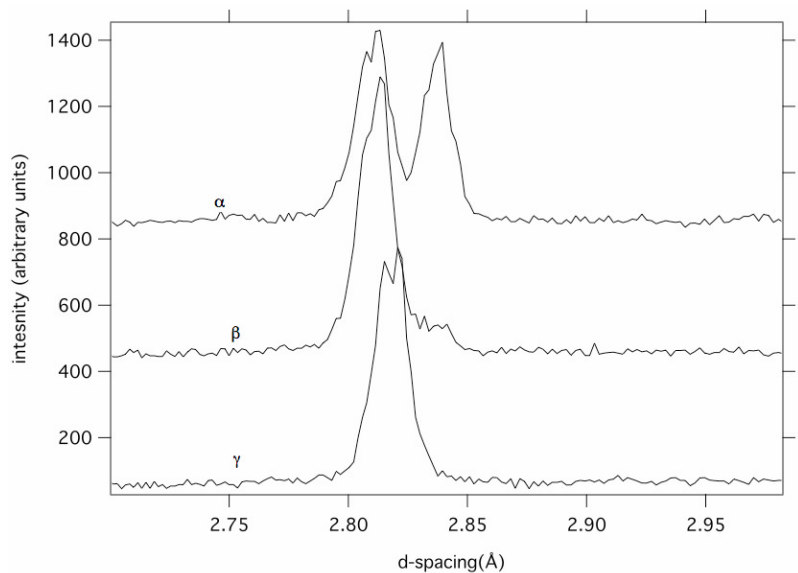

**Figure 1:** X-ray diffraction patterns of the (110)$_c$ peak of BiFeO$_3$ as a function of temperature through the $\alpha$–$\beta$–$\gamma$ phase transitions. Splitting in the rhombohedral ($\alpha$) and orthorhombic ($\beta$) phases disappears in the cubic $\gamma$-phase, where the slight peak asymmetry arises from the CuK$\alpha_1$/$\alpha_2$ splitting of the radiation used.

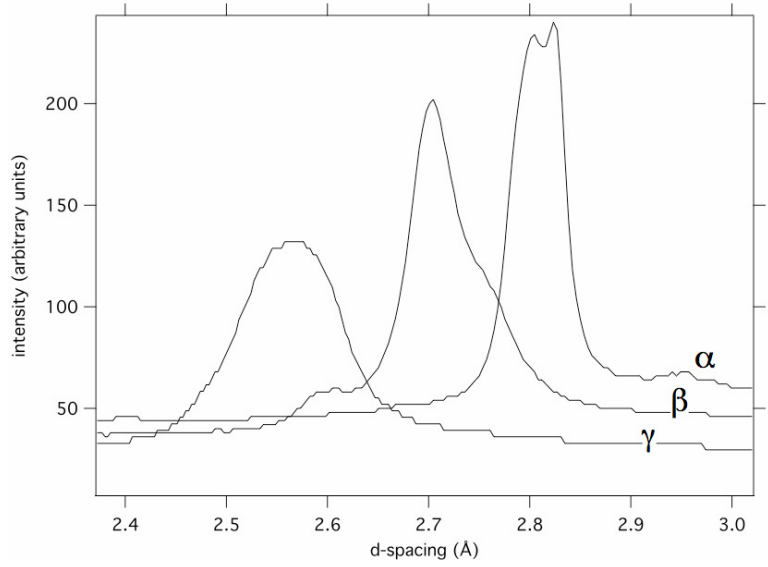

**Figure 2:** Resolved X-ray diffraction doublets of $(110)_c$ in $BiFeO_3$ as a function of pressure; above 47 GPa these become singlets. Although broadened, the cubic (110) of the $\gamma$-phase shows no sign of asymmetry.

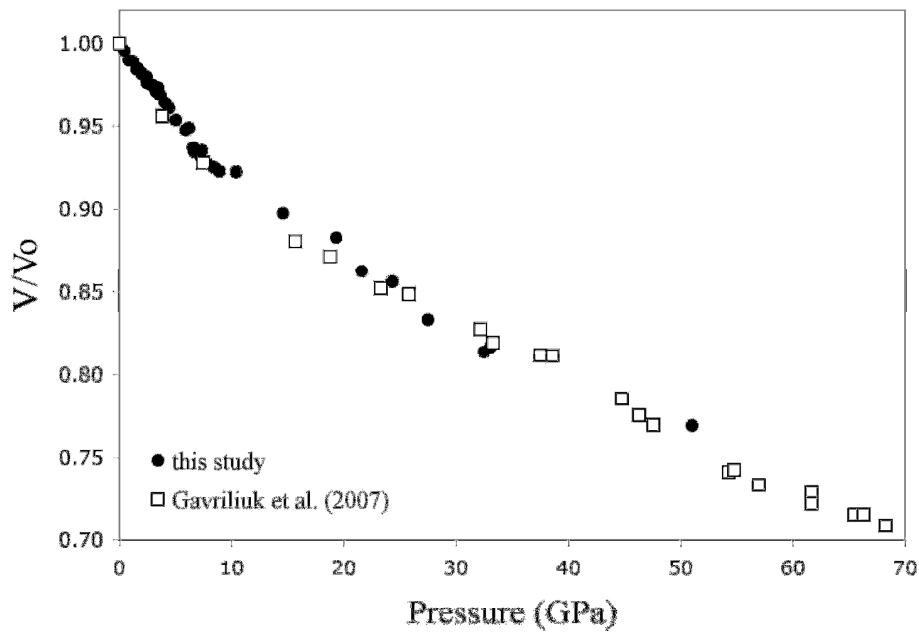

**Figure 3**: Unit cell volume as a function of pressure, both from the present study and from an earlier report by Gavriliuk et al [23]. The data at high pressures is consistent with a continuous evolution of the unit cell volume, indicative of a second order phase transition.

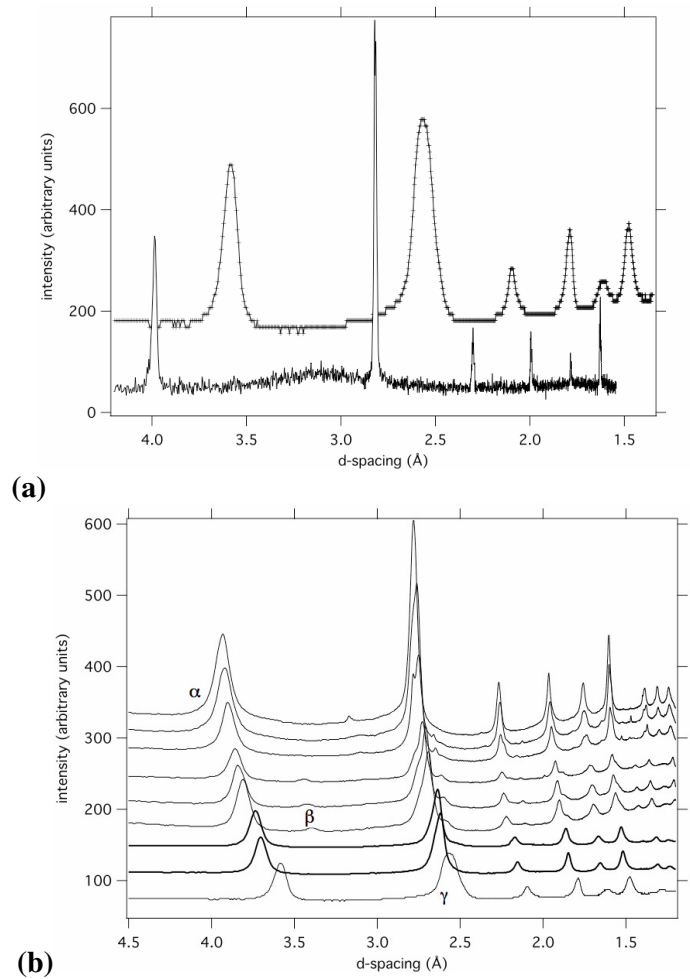

**Figure 4:** [a] Comparison of X-ray patterns of BiFeO$_3$ at high temperatures (below) and high pressures (above), showing the same cubic structure in the γ-phase with no indication of lower-symmetry super-lattice reflections. [b] Pressure evolution of the diffraction patterns of BiFeO$_3$ reveal the presence of the orthorhombic β-phase at intermediate pressures, indicated by the superlattice peak around 3.4 Å, which vanishes in the highest-pressure patterns of cubic γ-BiFeO$_3$.

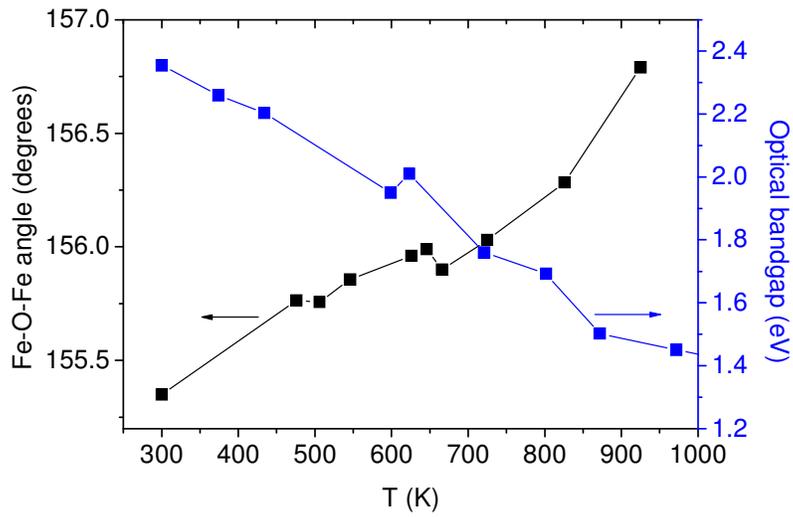

**Figure 5**: Fe-O-Fe bond angles (extracted from ref [39]) and optical bandgap (extracted from ref.[12] ) of BFO as a function of temperature. As the Fe-O-Fe bond becomes straighter, the bandgap decreases, consistent with increased orbital overlap between $Fe^{+3}$ and $O^{2-}$.